\documentstyle[fancybox,epsf,12pt,a4]{article}

\newcommand{\ta}{{\widetilde{a}}}

\newcommand{\tvarphi}{{\widetilde{\varphi}}}

\newcommand{\nn}{{\nonumber }}

\begin{document}

\begin{flushright}
 hep-th/0001044\\
 UT-838
\end{flushright}

\vfill
\begin{center}
{\LARGE Tachyon condensation and }\\ 
{\LARGE Boundary States in Bosonic String\footnote{
Talk given at YITP workshop ``Tachyon condensation and boundary state''
(October 27-29, 1999)}}\\
\vfill
{\large Y. Matsuo}\\
\vskip 10mm
{\large Department of Physics, University of Tokyo}\\
{\large Hongo 7-3-1, Bunkyo-ku, Tokyo 113-0034, Japan}
\end{center}
\vfill
\begin{center}
 {\bf Abstract}
\end{center}
\begin{flushleft}
 We discuss tachyon configuration for the unoriented
bosonic string theory which produces a bosonic string theory with
$SO(32)$ gauge symmetry in ten dimensions.
It is closely related to the tachyon condensation
scenario proposed by A. Sen.
We also give the boundary state description of the tachyon
condensation process, with some emphasis on the r\^ole of
orbifold conformal field theory.
\end{flushleft}
\vfill
\newpage

\section{Introduction}
After some intensive study carried out 
last year, (for example see \cite{r=Sen1}),
it becomes now evident that there are several motivations to 
study tachyon condensation scenario. For example,
(i) To seek correct vacuum of modular invariant but tachyonic
string theories, for example in bosonic strings,
type 0 (OA, OB), and also some heterotic strings.
(ii) To find stable non-SUSY soliton. Particularly successful
examples are the construction of the type I spinor particle
\cite{r=Sen2}, and also non-SUSY string junction \cite{r=junction}.
(iii) To find unstable anomaly free string theories \cite{r=typeK}.
Some examples are  $U(N)\times U(N)$ type IIB superstring
theory, $U(N)$ type IIA, and type I  analogue.
This is closely related to Witten's K-theory argument\cite{r=K}
since in his approach arbitrary D-brane is constructed
out of several pairs of $D$-9 and ${\bar D}$-9 branes.
Such a $D$-9 configuration gives rise to the extra
gauge symmetry in those theories.

Tachyon condensation is a dynamical process like Higgs mechanism.
At this moment, although there are many important references
\cite{r=boundary},
there are still many misteries in our understanding in terms
of conformal field theory. In this paper, we choose a particulary
simple example (bosonic string) and demonstrate the deformation
in terms of boundary conformal field theory explicitly.
We point out that the duality in conformal field theories
play essential r\^ole to understand the deformation.

\section{A brief review of tachyon condensation}

Consider $D$-$p$-- $\bar{D}$-$p$ system.
Open string sectors are labeled by $U(2)$ CP factor.
Fermion number operator for CP factor becomes nontrivial.
$$
(-1)^F_{CP} = \left(\begin{array}{cc}
	       1& 0 \\ 0 & -1
		    \end{array}\right)
 (=\sigma_3)
$$
For the total open string wave function 
$\Psi^{total}=\Psi^{osc}\otimes \Psi^{CP}$, the fermion number parity 
operator acts as,
$$
(-1)^F \Psi^{total}=((-1)^F_{osc}\Psi^{osc})\otimes
(\sigma_3\Psi^{CP}\sigma_3).
$$
It implies that:

(i) In $\Psi_{CP}=\sigma_0(=1),\sigma_3$ sector, we
have the usual GSO projection. There are
$U(1)_{\sigma_0}\times U(1)_{\sigma_3}$ gauge fields
at the massless level.

(ii) In $\Psi_{CP}=\sigma_1,\sigma_2$ sector, 
we have the opposite GSO projection which gives 
a complex tachyon field ($T, \bar{T}$).

We remark that  $T$ (resp. $\bar{T}$) have charge 2 (-2) 
under $A_{\sigma_3}$ but neutral under $A_{\sigma_0}$.
We assume that  the tachyon field has the following potential 
which is analogous to that of  Higgs fields. Based on this potential
we may discuss the process of the pair annihilation
of the D-branes in the similar fashion with 
more familiar symmetry breaking mechanism\footnote{
Recently there are some development to prove it by using the string 
field theory \cite{r=SFT}.}.

\centerline{\epsfbox{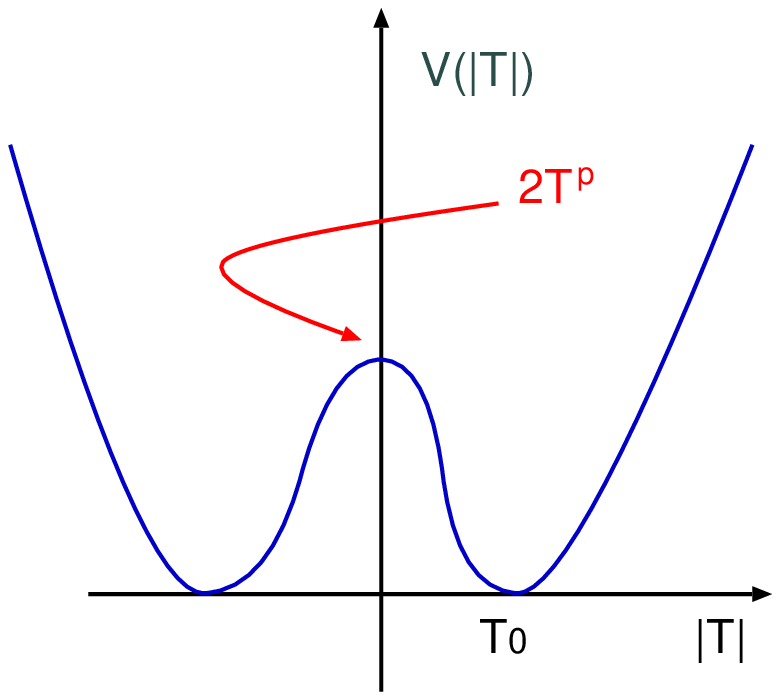}}

There are two equivalent scenarios to realize the
tachyon condensation. Let us start from $D$-$(p+2)$
and $\bar{D}$-$(p+2)$ branes.

\begin{itemize}
\item Step by step method proposed by A. Sen \cite{r=Sen1}:
	\begin{enumerate}
	\item First consider kink configuration for complex tachyon
	($T(x) \rightarrow \pm T_0$ as $x\rightarrow \pm \infty$)
	which define D-$(p+1)$ brane.
	\item Since moduli of tachyon is defined by $S^1$, such
	configuration is topologically unstable.
	\item On this D-$(p+1)$ brane, there is a real-valued tachyon
	field. One may again consider a kink configuration,
	which defines $D$-$p$ brane.
	\item Since the moduli is $Z_2$, this tachyon configuration
	becomes topologically stable which defines $D$-$p$ brane.
	\end{enumerate}

\item Topological method proposed by E. Witten \cite{r=K}:\\
We consider a topologically nontrivial configuration
(vortex) for gauge fields and tachyons in two dimensions,
\begin{eqnarray}
 \frac{1}{2\pi}\int F_{\sigma_3} & = & n \qquad \frac{1}{2\pi}\int F_{\sigma_0} =  0\nonumber\\
 T(x) & \sim & e^{in\theta} |T_0| \qquad \mbox{as}\quad |x|\rightarrow \infty
  \nonumber\\
&\sim& x_1\sigma_1 + x_2 \sigma_2 + O(|x|^2).\nonumber
\end{eqnarray}
D-$p$ and 
D-${\bar p}$ can be  pair annihilated if tachyon field takes $|T|=|T_0|$
at spacial infinity. 
Then n D-$(p-2)$ branes appear at the core of vorticity. 
$U(1)_{\sigma_0}\times
U(1)_{\sigma_3}$ is broken to $U(1)_{\sigma_0}$.\footnote{
Fate of $U(1)_{\sigma_0}$ is known to be Witten's U(1) problem.
See for example \cite{r=U1} for the attempts to understand the issue.}
\end{itemize}

BCFT method is easier to be applied to the first scenario.
On the other hand, its topological nature is illuminated
in the second. We will use both of them.

\section{A Scenario for Bosonic string}

In the following sections, we will mainly consider
the tachyon condensation of the bosonic string by following 
\cite{r=Matsuo}.
Some of the key motivations are
(i) it is the simplest string models, and
(ii) it is nonetheless the most generic string theory.
It is supposed to contain superstring theories, type 0 theories,
heterotic strings as its particular vacuum.
It is therefore important  to seek the fate 
of the bosonic string if any.
It would be also desirable if one may find a
dynamical scenario for the compactification of
the 26 dimensional space-time.

A possible candidate of the destination of
the open bosonic string would be,
\begin{enumerate}
 \item $SO(32)$ type I string
 \item Type 0 string ($SO(32)\times SO(32)$ theory) which
is more plausible since it 
has bosonic spectrum and
 still contains a closed string tachyon.
\end{enumerate}

Open unoriented bosonic string 
was studied by some groups \cite{r=bosonic},
and it has been known that a consistent model exists with
$D=26$ bosonic string and space filling $D$-25-branes.
Tadpole free condition restricts the number of
25-branes to be $2^{13}=8192$. This condition comes
from the cancellation of massless excitation
arising from the boundary states of the open boundary
and the crosscap.
If this condition is satisfied, the model becomes a finite theory
if one applies the zeta function regularization to the
infinity from the tachyon mode.

Although it is very interesting that such a consistent model exists
in purely bosonic theory, it is obviously discouraging that
\begin{enumerate}
 \item it has a huge gauge symmetry $SO(8192)$ to be realistic,
 \item it is defined only in 26 dimensions,
 \item it has many open string tachyons since there
is no GSO projection,
 \item there is no D-brane charges and every D-branes becomes unstable.
\end{enumerate}
In the following discussions, we would like to point out that
these weak points may be eliminated by using a
topologically nontrivial configuration for the tachyon fields.

\subsection{Sen's argument}

Sen \cite{r=Sen3} applied applied his step by step method to the system with 
two D-$(p+1)$-branes. There are four tachyons arising
from four CP sectors of  the open string. One may introduce
$Z_2$ Wilson line to
one of the D-branes to make the intertwining open string
anti-periodic in one space direction.
One D-$p$-brane will appear at the position of the 
kink of the tachyon field,\footnote{
Exact description in terms of boundary state will be 
discussed in the next section.} while original two D-branes
disappear.  In the exact treatment of such process,
one need to compactify one of the space direction on a circle.
By such compactification, the mass squared of tachyon field
increases while the radius of $S^1$ become smaller.
At a specific radius, the tachyon mode becomes massless
and the exact treatment of the tachyon condensation becomes possible.
In this process, one is forced to
reduce the number of the uncompactified dimensions.
Successive application of this idea to $SO(2^{13})$ theory
would imply that one $D$-12 brane will be produced at the
end of tachyon condensation while we have
13 space-time dimensions. Since such a theory can not be
tadpole free in any respect, this scenario seems rather unnatural.

\centerline{\epsfbox{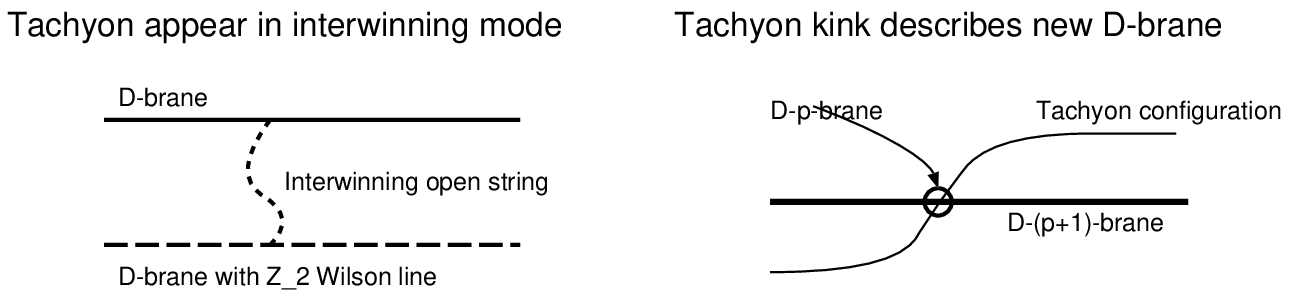}}

Sen have also argued that there are other options.
\begin{itemize}
 \item Starting from a single D-brane, there exists an open string
       tachyon. If one consider the kink for this tachyon mode,
       one D-$(p+1)$-brane may produce several D-$p$-branes.
 \item If tachyon develops several kinks, there may appear several
       D-$p$-branes. Therefore, the number of D-branes can be arbitrary.
\end{itemize}
The first possibility seems not plausible since Cardy's constraint
\cite{r=Cardy}
on the boundary state implies that it is not possible to have
an anti-periodic tachyon field in case of a single D-brane.
The second possibility  seems rather artificial.
Such an arbitrariness always arises if we seek
the tachyon condensation in topologically unstable configurations.
However,  as wee see, one may achieve some kind of 
topological stability even in the case of bosonic string.
If we restrict our scope to the topologically stable sectors,
there is no such arbitrariness.

\subsection{Our proposal}

It seems to us that the only topologically 
stable tachyon condensation scenario is  a process
where two D-$(p+2)$-branes with tachyon condensation
produce a single D-$p$ brane while compactifying two
space dimensions toroidally. In this scenario, we have
$2^{13-d/2}$ D-branes while compactifying $d$ dimensions.
It matches with the tadpole free conditions
for generic open string. If we take $d=16$, we have
$SO(32)$ theory at dimension 10 which is a candidate
model which can be dual to type I or type 0 string theories.

To explain the topological nature, let us consider two 
D-branes. The gauge group associated with it
is $SU(2)$ except for the overall $U(1)$. 
The vacuum moduli of the tachyon potential is $S^3$.
In such a situation, one may
not expect topologically stable configuration since
$\pi_1(S^3)=0$.

However, at the special infinity, the gauge transformation
can be nontrivial. We recall that the
open string transforms in adjoint representation.
The actual gauge group which acts nontrivially
on the open string is not $SU(2)$ but $SO(3)$.
In this context one may have a topologically 
nontrivial loop in the gauge configuration space since, 
$$\pi_1(SO(3))={\bf Z}_2.$$
It also means that the topologically nontrivial sector
is in a sense unique.  One may interpret it that the number of the 
D-$p$ brane which will appear at the core is restricted to be 
one.

To illustrate this idea more explicitly, we compactify two dimensions
toroidally and require the behavior of the tachyon field at the core 
of vorticity,
\begin{equation}
T(x) \sim x^1 \sigma^1 + x^2\sigma^2\quad
(\mbox{as } |x|\rightarrow 0).
\end{equation}
Tachyon must be anti-periodic in $x^i$ directions
in $\sigma^i$ sector. This configuration appeared in
\cite{r=K} to describe nontrivial vortex configuration.

Roughly speaking, such a configuration is realized if we introduce 
a kind of non-abelian Wilson lines which produces curvature,
$$
A_1 \propto \sigma_2\quad
A_2 \propto \sigma_1
$$
It costs some energy but since they turn out to be in the 
topologically nontrivial, such a configuration is stable.
A simple realization is possible if we impose the following
boundary condition for the open string,
\begin{eqnarray}
 \Psi(x^1+2\pi R, x^2) & = & \sigma_2 \Psi(x^1,x^2)\sigma_2,\nonumber\\
 \Psi(x^1, x^2+2\pi R) & = & \sigma_1 \Psi(x^1,x^2)\sigma_1.\nonumber
\end{eqnarray}
It belongs to the topologically nontrivial sector since
the gauge transformation (=Wilson line) around the loop is 
$\sigma^1\sigma^2(\sigma^1)^{-1}(\sigma^2)^{-1}=-1$
\footnote{This is a projective representation of the lattice
translation group. Similar nontrivial representation in CP appeared in
the description of the discrete torsion in orbifold conformal
field theory\cite{r=DT}.}.
We summarize the periodicity along each direction,
\begin{center}
\begin{tabular}[t]{@{\vrule width 1pt}c||c|c|c|c@{\ \vrule width 1pt}}
\hline
 &$\sigma^0$ &$\sigma^1$ &$\sigma^2$ & $\sigma^3$\\ \hline\hline
$x^1$ & + & $-$ & + & $-$\\
$x^2$ & + & + & $-$ & $-$\\ \hline
\end{tabular}
\end{center}
Gauge symmetry is broken from $U(2)$  to $U(1)$ which
is quite unusual but necessary to describe pair annihilation of
two D-branes.

Since we simply impose the antiperiodic boundary condition
to some components of the open string, the spectrum of
the topologically nontrivial sector is very easy to calculate.
If the original open string lives in radius $R$, the momentum
distribution of the twisted theory is the equally separated
momenta with the separation $1/2R$. Therefore, 
the spectrum of the topologically nontrivial sector
becomes identical the either of the following two systems,
\begin{enumerate}
 \item one D-$(p+2)$ brane in radius $2R$,
 \item one D-$p$ brane in radius $1/2R$.
\end{enumerate}
It supports our expectation that only one D-$p$ brane
appear at the core of vorticity.

To define the unoriented string theory, we need to
impose the orientation projection $\Omega$ to the open string
wave function. At each mass level, we have project out
either symmetric (anti-symmetric) part of the wave function.
Such the projection is compatible with our boundary
condition since Pauli matrices satisfies well-defined
parity under adjoint action of $\Omega$.
The symmetry breaking pattern $U(2)\rightarrow U(1))$
remains the same but the precise correspondence
in the  momentum distribution is corrupted.

To extend our analysis to the compactification of the higher
dimensional tori is straightforward.
To have nontrivial configuration in $2d$ dimensions, general
prescription will be,
\begin{eqnarray}
&&\Psi(x^{26-2d},\cdots,x^{25-2d+i}+2\pi R,\cdots,x^{25})\nonumber\\
&&\quad= \tilde\Gamma_i \Psi(x^{26-2d},\cdots,x^{25-2d+i},\cdots,x^{25})
\tilde\Gamma_i\nonumber
\end{eqnarray}
where $\tilde \Gamma$ satisfies,
\begin{eqnarray}
&&\left\{\tilde\Gamma_i,\tilde\Gamma_j\right\}= 2\delta_{ij}\nonumber\\
&&\left\{\tilde\Gamma_i,\Gamma_i\right\}=0\quad
\left[\tilde\Gamma_i,\Gamma_j\right]=0 \quad (i\neq j)\nonumber
\end{eqnarray}
With this prescription, $\Psi$ behaves at the core as,
$$
\Psi(x)=\sum_{i=1}^{2d} x^{25-2d+i} \Gamma_i + O(|x|^2)
$$
which is identical to the tachyon configuration of
the superstring as discussed by Witten \cite{r=K}.
With this boundary condition, Chan-Paton gauge
symmetry is broken from $U(2^d)$ to $U(1)$.

\subsection{Symmetry enhancement from closed strings?}

At $R=1/\sqrt{2}$, we get extra gauge symmetry from
vertex operator $e^{i X(z)/R}$ in the closed string sector.
In general, if the compactified direction is given by the root lattice
of $G$, gauge group is enhanced to $G$.
If we compactify $16$ dimensions, the maximal enhancement
is given by $SO(32)$ or $E_8\times E_8$.
This well-known mechanism was used to construct the heterotic
string theory.

Tachyon fields which interpolates the different D-branes
becomes anti-periodic in the compactified direction
and it has  the mode expansion,
$$
T(x) = \sum_{n\in {\bf Z}} T_{n-1/2} e^{i(n-1/2)X}.
$$
Mass of each mode $T_{n-1/2}$ is given by $m^2=-1+(n-1/2)^2/2R^2$.
The lowest mode $T_{\pm 1/2}$ becomes massless when $R=1/2\sqrt{2}$.
We expect that the existence of massless mode signifies the
enhanced gauge symmetry.  However the radius for the massless
tachyon is half of that of the closed string gauge enhancement
point.

\section{Tachyon condensation and Boundary CFT}

As we saw in the last section, if we fix the radius
of the target space to $R=1/2\sqrt{2}$, some of the tachyon mode
becomes massless. In the case of the closed string excitation,
such a massless mode is used to
deform the background (metric, antisymmetric tensors etc.) 
of the target space. On the other hand, the deformation
by the massless open string mode
usually triggers the deformation of the
{\em D-brane} to which open string is attached.
In our situation, the deformation by massless mode
induces the tachyon condensation and we want to follow
the change of the D-brane state in such a process
as exactly as possible.
In this section, we give an explicit description
of boundary conformal field theory in Sen's scenario.

Before we discuss the detail, it is useful to
to indicate that the radius $R=1/2\sqrt 2$ is
exactly the point where the moduli space of toroidal compactification
and orbifold compactification meets. 
This point is quite essential in the following discussion.

\centerline{\epsfbox{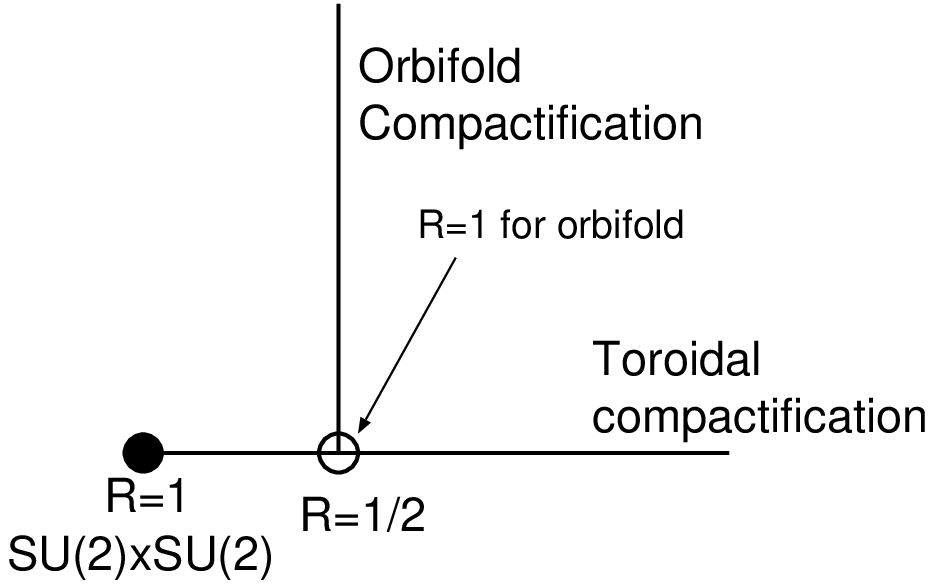}}

\subsection{Analysis in open string Hilbert space}

In this section, we review Sen's discussion \cite{r=Sen3}
of the analysis which uses the open string Hilbert space.
To treat the condensation of the tachyon field, Sen introduced 
new variables to describe the compactified direction $X$,
\begin{eqnarray}
 e^{i\sqrt{2}X} & = & \partial Y + i \partial Z\nonumber\\
 e^{i\sqrt{2}Y} & = & \partial X - i \partial Z\nonumber\\
 e^{i\sqrt{2}Z} & = & \partial X + i \partial Y\nonumber
\end{eqnarray}
By using the new variables, tachyon condensation is equivalent to
the deformation induced by  $\oint \partial Y \otimes \sigma_1 dz$.

As we emphasized before, the radius for the massless
tachyon is not equal to the radius for the gauge enhancement
in the closed string sector.
Although we introduced the variables of $SU(2)$ current algebra,
we need to project out some part of the spectrum.
We should not forget that $e^{\pm i \sqrt{2}X}$ is
antiperiodic in direction $X$.

Such a projection can be carried out
by introducing two parity operators,
\begin{eqnarray}
 h & \equiv &  X \rightarrow X +\pi/\sqrt{2}\nonumber\\
 g & \equiv &  X \rightarrow -X\nonumber
\end{eqnarray}
The first one is the translation in $X$ direction. Open string
modes in $\sigma^1, \sigma^2$ sectors becomes odd under this 
operator. 

On the other hand, we do not originally have any projection of
the second parity operator. However, to define
a similar projection in $Y$ variable,
these two operators change their r\^ole,
$$
h_X = g_Y \qquad g_X = h_Y .
$$
This relation motivates us to introduce the second operator.

Let us now proceed to define necessary projections
in terms of $Y$ variable.
Originally in $Y$ direction, we have both boundary condition
$$
\Psi(Y+2\pi) \equiv g \Psi(Y) = \pm \Psi(Y).
$$
Since we have both periodic and anti-periodic boundary conditions,
it is more natural to extend periodicity in $4\pi$.
$$\Psi(Y+4\pi)=\Psi(Y)$$

By the tachyon condensation deformation in  $Y$ direction,
we deform this boundary condition to
\begin{eqnarray}
 \Psi(Y+4\pi)&=& \sigma_1 \Psi(Y)\sigma_1\nonumber
\end{eqnarray}

The eigenvalues of $h$ and $g$ in each sectors are now deformed.
We summarize it in the next table,
\begin{center}
\begin{tabular}[t]{@{\vrule width 1pt}c||c|c@{\ \vrule width 1pt}}
\hline
 &$h_{Y}
=g_{X}$ &$g_{Y}=h_{X}$ 
\\ \hline\hline
\ \ $I$ & $\pm 1$ & $+$ \\
\ \ $\sigma_1$ & $\pm 1$ & $-$ \\
\ \ $\sigma_2$ & $\pm i$ & $-$ \\
\ \ $\sigma_3$ & $\pm i$ & $+$ \\ \hline
\end{tabular}
\end{center}

If we combine the four sectors, we have both signs
in $g_Y$ parity. For each sign, we have four twists $\pm 1, \pm i$
in the boundary condition.
This indicates that momentum is now
quantized in the unit $1/4$ i.e. the compactification radius
is apparently changed to $\sqrt 2$.
If we take the T-dual in $Y$  direction,
Radius becomes $1/2\sqrt{2}$ --- original radius 
and two D $(p+1)$-branes become single $p$-brane.

\subsection{Tachyon condensation in Boundary state}

We now turn back to describe this tachyon condensation process
in terms of the boundary state.
The deformation is induced by insertions of 
$$\exp\left( i \varphi\oint dz \partial Y \otimes\sigma_1\right)$$
at the boundary.

\centerline{\epsfbox{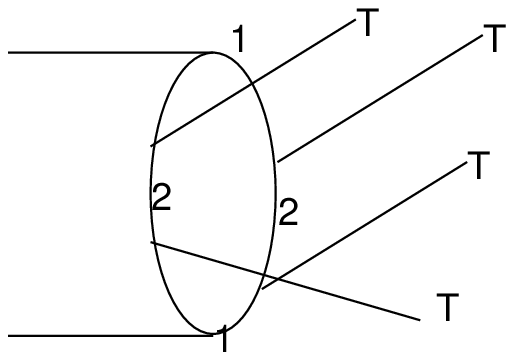}}

There are two difficult points which we need some care
to discuss the deformation.

The first one is that the D-brane at the boundary 
must be switched by CP factor $\sigma_1$ at each insertion point
as illustrated in the above figure.
At each end of the inserted open string, we have different
D-branes and we need to toggle CP factor in two ends.

The insertion of off-diagonal CP factor is usually treated by
using the trace.   In our case, we need some care
since we have already deformed one of the D-brane by using
$Z_2$ Wilson line.

The second point is that the deformation operator
is the vertex operator in terms of original field,
$\partial Y=\cos(X)$. 
We have to change the dynamical variable to $Y$
as in the open string approach.
The $Z_2$ transformation $h$ act on $Y$
as $Y\rightarrow -Y$. 
It suggests that the natural framework is to
use $Z_2$ orbifold variable.

To describe orbifold CFT, it is useful to
prepare the boundary states
for the theory with $S^1$ compactification.
We have two types of boundary states,
\begin{itemize}
 \item Dirichlet:
$|D(\varphi)\rangle$ \hskip 10mm ($\varphi\sim \varphi + 2\pi r$)\\
$\frac{1}{\sqrt{2r}}\sum_{k=0}^\infty
e^{-ik\varphi/r}\exp\left(-\sum_{n=1}^\infty a_n^\dagger
\ta_n^\dagger\right)|0,k\rangle$
 \item Neumann:$|N(\tvarphi)\rangle$
\hskip 10mm ($\tvarphi\sim \tvarphi + \pi/r$) \\
$ {\sqrt{r}}\sum_{w=0}^\infty
e^{-2irw\tvarphi}\exp\left(\sum_{n=1}^\infty a_n^\dagger
\ta_n^\dagger\right)|w,0\rangle$
\end{itemize}
Physical interpretation of the deformation parameters
are well-known to be
location of D-brane for $\varphi$ and the Wilson line
for $\tvarphi$.
In this language, the boundary state before tachyon condensation is
\begin{equation}\label{e=BS1}
|N(0)\rangle + |N(\pi/2r)\rangle.
\end{equation}

The boundary states for $S^1/Z_2$ orbifold
was discussed in \cite{r=OA}.

\centerline{\epsfbox{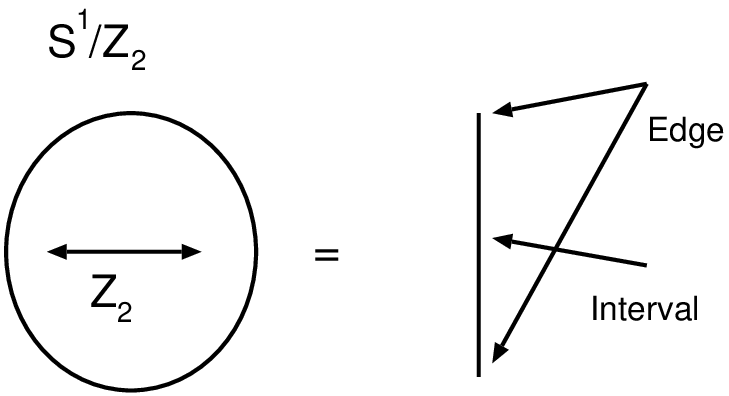}}

In this case, the boundary state which
describes the location of D-brane in the interval 
and those located at the edges are essentially different.
We write them by using the boundary states for 
$S^1$ compactification.

1. Boundary state for interval \\
($0<\varphi<\pi r$, $0<\tvarphi<\pi/2 r$ )
\begin{eqnarray}
|D_O(\varphi)\rangle& =& \frac{1}{\sqrt 2}
(|D(\varphi)\rangle+|D(-\varphi)\rangle)\nn\\
|N_O(\varphi)\rangle &=& \frac{1}{\sqrt 2}
(|N(\varphi)\rangle+|N(-\varphi)\rangle)\nn
\end{eqnarray}

2. Boundary state for edges (half D-branes)
($\varphi=0,\pi r$, $\tvarphi=0,\pi/2r$):
\begin{eqnarray}
|D_O(\varphi)\pm\rangle&=& \frac{1}{\sqrt 2}|D(\varphi)\rangle
\pm \frac{1}{2^{1/4}} |D(\varphi)_T\rangle\nn\\
|N_O(\tvarphi)\pm\rangle&=& \frac{1}{\sqrt 2}|N(\tvarphi)\rangle
\pm \frac{1}{2^{1/4}} |N(\tvarphi)_T\rangle\nn
\end{eqnarray}

Open string partition function with specific $Z_2\times Z_2$ charge
(generated by $h$ and $g$)
can be written in terms of edge boundary states:

\begin{eqnarray}
 Z_{++} & = & \frac{1}{2}\frac{\sum_n q^{4n^2}}{\prod_n(1-q^n)}
  +\frac{1}{2}\frac{1}{\prod_n{(1+q^n)}}
 = \langle N_O(\tvarphi_\pm)\pm |\Delta|N_O(\tvarphi_\pm)\pm \rangle\nonumber\\
 Z_{+-} & = & \frac{1}{2}\frac{\sum_n q^{4n^2}}{\prod_n(1-q^n)}
  -\frac{1}{2}\frac{1}{\prod_n{(1+q^n)}}
 = \langle N_O(\tvarphi_\pm)\pm |\Delta| N_O(\tvarphi_\pm)\mp \rangle\nonumber\\
 Z_{-+} & = & \frac{1}{2}\frac{\sum_n q^{4(n+1/2)^2}}{\prod_n(1-q^n)}
 = \langle N_O(\tvarphi_\pm)\pm |\Delta|
 N_O(\tvarphi_\mp)\pm \rangle\nonumber\\
 Z_{--} & = & \frac{1}{2}\frac{\sum_n q^{4(n+1/2)^2}}{\prod_n(1-q^n)}
 = \langle N_O(\tvarphi_\pm)\pm |\Delta|
 N_O(\tvarphi_\mp)\mp \rangle\nonumber
\end{eqnarray}
where $\tvarphi_\pm=0,\pi/2r$.

By choosing specific radius $r=1/2\sqrt{2}$, two $Z_2$ symmetry
becomes interchangeable,
$$
Z_{+-}=Z_{-+}.
$$
This accidental symmetry is a consequence of the fact that
this model is an orbifold CFT and at the same time 
a CFT on $S^1$.
In terms of edge boundary states, interchange of
$h$ and $g$, (or $J_3$ and $J_1$) is described by
the change of the r\^ole of two edges and the sign of the twisted sector,
$
| N_O(\tvarphi_{\epsilon_1})\epsilon_2 \rangle
\leftrightarrow
| N_O(\tvarphi_{\epsilon_2})\epsilon_1 \rangle.
$

To summarize, the deformation associated with
the tachyon condensation can be achieved by the following steps.
(i) Prepare the boundary state which describe two bosonic
D-branes with $Z_2$ Wilson line on one of them (\ref{e=BS1}).
(ii) Reinterpret the boundary state as those of $S_1/Z_2$
orbifold model.
(iii) Change the r\^ole of $J_1$ and $J_3$.  In terms of the
boundary state, it amounts to the interchange of 
two parameters $\tilde\varphi$ and $\epsilon$.
(iv) Deform the parameter $\tilde\varphi$
of one of the D-brane boundary states.
More explicitly, it can be carried out as follows.
\begin{eqnarray}
|N(0)\rangle + |N(\pi/2r)\rangle 
&=&   \sqrt{2}\sum_{\epsilon_1,\epsilon_2}
 |N_O(\tvarphi_{\epsilon_1})\epsilon_2\rangle\nn\\
&=& \sqrt{2}  \sum_{\epsilon_1,\epsilon_2}
 |N_O(\tvarphi_{\epsilon_2})\epsilon_1\rangle\nn\\
& \rightarrow & \frac{1}{\sqrt{2}}\sum_{\epsilon_1,\epsilon_2}
 |N_O(\tvarphi_{\epsilon_2})\epsilon_1\rangle
+ {\sqrt{2}} |N_O(\frac{\pi}{\sqrt{2}})\rangle\nn
\end{eqnarray}
We note that the boundary states
in the final line  produces the same
partition function in the open string sector as that
of $|D(0)\rangle$ for $S^1$ compactified model.
This calculation explicitly demonstrates 
the production of one D-$p$ brane from two D-$(p+1)$
branes by the tachyon condensation.
However, we have to mention one subtlety.
In passing from the second to the third line,
the intermediate state in the course of deformation
can not be properly interpreted as D-brane boundary
state since it would produce partition function
with fractional coefficient  in the open string sector.
Namely, it does not satisfy Cardy's constraint.
In this sense, one can not describe the tachyon condensation
as the continuous deformation of the boundary state.
This phenomena was also discussed by Sen and he
explained it by showing the appearance tadpole.
In this sense, it is more appropriate that two vacua (a) two D-$p+1$ brane
system and (b) one D-$p$ brane system are not connected continuously
but belong to the different topological sectors.


\section{Discussion}

In the second half of this paper, I discussed 
duality with orbifold theory is essential to describe
tachyon condensation.
This seems to be a generic feature since massless
tachyon is always described by vertex operator
with anti-periodic boundary condition.
In our analysis, the twisted sectors actually do not
play any r\^ole since we always sum over various sectors.
 I hope that these states may have some meanings to
understand still mysterious Witten's $U(1)$ problem
In superstring case, $Z_2$ operator $g$ is replaced
 by $(-1)^F$. Orbifold CFT is described by KT point.



\begin{thebibliography}{99}

\bibitem{r=Sen1} A.~Sen, hep-th/9904207 and references therein.
\bibitem{r=Sen2} A.~Sen, hep-th/9808141.
\bibitem{r=junction} A. ~Sen and B. Zwiebach, hep-th/9907164.
\bibitem{r=typeK} M. Srednichi, JHEP 9808 (1998) 005,
S. Sugimoto, hep-th/9905159.
\bibitem{r=K} E. Witten, hep-th/9810188;\\
 P. Horava, hep-th/9812135.
\bibitem{r=boundary} O. Bergman, H. Gaberdiel, hep-th/9806155;\\
O. Bergman, H. Gaberdiel, hep-th/9908126;\\
M. Frau, L. Gallot, A. Lerda, P. Strigazzi, hep-th/9903123;\\
P. Di Vecchia, Lecture given in this workshop, hep-th/9912275.
\bibitem{r=SFT} A. Sen, hep-th/9911116;\\
A. Sen and B. Zwiebach, hep-th/9912249.
\bibitem{r=U1} P. Yi, hep-th/9901159.
\bibitem{r=Matsuo} Y. Matsuo, hep-th/9905159.
\bibitem{r=bosonic} S. Weinberg, Phys. Lett. 187B (1987) 278;\\
M. Douglas and B. Grinstein, Phys. Lett. 183B (1987) 52;\\
N. Marcus and  A. Sagnotti, Phys. Lett. 188B (1987) 58.
\bibitem{r=Sen3} A. Sen, hep-th/9902105.
\bibitem{r=Cardy} J. L. Cardy, Nucl. Phys. B324 (1989) 581.
\bibitem{r=DT} M. Douglas, hep-th/9807235;\\
M. Douglas and B. Fiol, hep-th/9903031.
\bibitem{r=OA} M. Oshikawa, I. Affleck, cond-mat/9612187.

\end{thebibliography}
\end{document}